# Non-recursive Approach for Sort-Merge Join Operation

Norah Asiri(B) and Rasha Alsulim

Computer and Information Sciences College,
King Saud University, Riyadh, Saudi Arabia asiri.norah.m@gmail.com,
r.m.z1433@hotmail.com

**Abstract.** Several algorithms have been developed over the years to perform join operation which is executed frequently and affects the efficiency of the database system. Some of these efforts prove that join performance mainly depends on the sequences of execution of relations in addition to the hardware architecture. In this paper, we present a method that processes a many-to-many multi join operation by using a non-recursive reverse polish notation tree for sort-merge join. Precisely, this paper sheds more light on main memory join operation of two types of sort-merge join sequences: sequential join sequences (linear tree) and general join sequences (wide bushy tree, also known as composite inner) and also tests their performance and functionality. We will also provide the algorithm of the proposed system that shows the implementation steps.

· Concurrent operations

## 1 Introduction

Database query operations facilitate the ease of information retrieval from one or more relations. However, binary join operation is one of the most challenging operations to be implemented efficiently. It is the only relational algebra operation that combines the related tuples from different relations among different attributes schemes [10].

The improvement of the performance of any database system necessitates improvement of the frequently executed operations such as join because it depends on transferring and moving data to/from the main memory [10]. Thus, the optimization for this process should be offered to reduce its expenses and to improve its functionality. The join operation which is denoted by  is used to combine related tuples from two relations into a single longer tuple [11].

The join operation over two datasets R and S with binary predicate t and attributes a and b is given as:

$$R \bowtie S = \{t | t = rs \land r \in R \land s \in S \land t(a) = t(b)\} \tag{1}$$

There are many types of relationships among relations, such as one-to-one, one-to-many and many-to-many. A many-to-many relationship refers to a relationship between tables in a database when a parent row in one table contains several child rows in the second table and vice versa. Currently, many-to-many relationship is usually a mirror of the real-life relationship between the objects that the two tables represent. In this paper, we investigate sort-merge with multi-join queries and try to optimize the join process and to achieve efficient execution [5,13].

The rest of this paper is organized as follows: in Sect.2, background of join strategies. In Sect.3 methodology that are used in our approach. Our experiment and results are given in Sect.4. Finally, conclusions from our findings are presented in Sect.5.

## 2   Background

### 2.1   Join Strategies

The strategies used to perform join operations are discussed in this section, were some are implemented while others are just proposed. The debate over which are the comparative performance join algorithms of these approaches has been going on for decades. The main reason behind discussing these methods is to identify which one is the best and most applicable with our suggested system.

1. **Simple Nested-Loops:** is considered to be the simplest form of join. It starts from the inner loop which is designated to the inner relation and the other loop which is tied to the outer relation. For each tuple in the outer relation, all tuples in the inner relation are scanned and compared with the current tuple in the outer relation. In case of matching the condition of join, the two tuples are joined and positioned in the outer buffer [10]. In practical view, nested-loop join is performed as a nested-block join, because the tuples are retrieved in form of blocks rather than individual tuples [11].
   In this algorithm, the total amount of reduction in I/O activity depends on the size of the available main memory; it is noticed that each tuple of the inner relation is compared with every tuple of the outer relation. In this way, the execution of this algorithm requires $O(n \times m)$ time for joins execution.

2. **Nested-Loops Join with Rocking:** to optimize the simple nested-loops join method, we can use an extra step to ensure it works with more efficiency. This step is called: rocking the inner relation [8]. Rocking the inner relation means to read the inner relation from top-most to bottom for one tuple only of the outer relation and from bottom to top for the next tuple. Consequently, the I/O overhead is reduced since the last returned page of the inner relation in the current loop is also used in the next loop.

Granting all this, the exhaustive matching strategies and the poor efficiency in both simple and rocking nested-loops join makes it inappropriate for joining large relations.

3. **Hash Join Methods:** a large number of join processes uses the hash methods. With the simple hash join methods, the join attribute(s) values at the inner relation are hashed using hash function to create a hash table with keys of inner relation and then sort them. After that the partitioning phase carried out by looping over the tuples of outer relation. For each key in the outer relation, search for matching keys in the hash table and attach the matching tuple(s) to the output. Its search execution is $O(1)$ which outperforms other join methods in some cases and makes it an accepted solution to researchers. However, it has some disadvantages such as the duplicate keys in inner relation which causes conflict in the hash table. Also, there are extra irrelevant comparisons resulting from the size of the hash table. Moreover, it has a limitation due to current CPU architecture [7].

4. **Sort-Merge Join:** relies on sorting the rows in both input tables by the join key and merges these relations. Sort-merge performance mainly depends on choosing an efficient sorting implementation where more than 98% of sortmerge method costs lies [7]. The investigative studies show that hash join needs at least 1.5X more memory bandwidth than sort merge join [7]. Furthermore, authors in [3] proves that in multi-core databases and modern multi-core servers, sort-based join beats hash-based parallel join algorithms and mostly has a linear relationship with the number of cores [3]. For the future computer system architectures, if the growth gap between compute and bandwidth continues to expand, the sort-merge join will be more effective than hash join [7].

## 2.2 Multi-join Queries (Linear and Bushy Trees for Many-to-Many Multi-join Queries)

Investigators have argued and worked on the implementation and the performance of parallel and concurrent DBMS. To optimize a multi join query response time, the exploitation of concurrency by using trees is used. Hence, any differ-ences in response time result from the differences in the shape of the tree [14]. This paper presents analytical experiments of two types of join sequences [12].

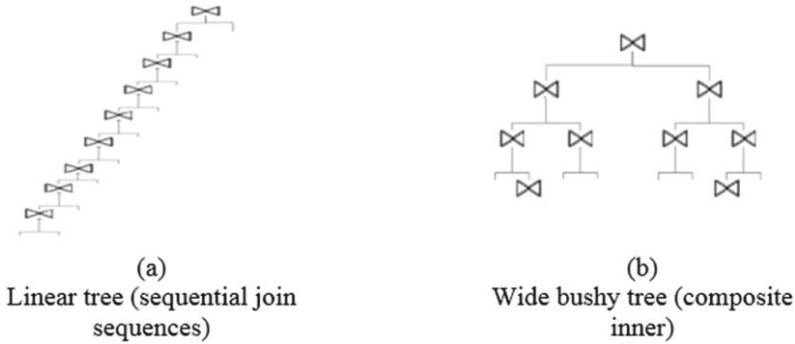

**Fig.1.** Linear and wide bushy trees multi-join sequences shapes

The first type is the linear tree or sequential join sequence, in which the resulting relation of an intermediate join can only be used in the next join. For instance, in Fig.1(a) where every non-leaf node (internal node) denotes the resulting relation from joining its child nodes. The second type is known as general join sequence (wide bushy tree, also known as composite inner) [6] in which the current resulting relation of a join is not mandatory to be only used in the next join as shown in Fig.1(b).

## 3    Methodology

Most references present tree traversal using recursion only. In [2], the literature survey shows that most references only indicate the implementations of the recursive algorithms, and only few references address the issue of non-recursive algorithms. In our investigation, we use a non-recursive algorithm that is simple, efficient which depends on a stack and a post-order binary tree traversal. We dynamically allocate the binary trees elements in a way that each element (node) has at most two potential successors.

We cover two kinds of binary trees: the wide bushy tree and the linear tree. Particular multi-join expressions are applied to the tree because all of the join operations are binary. It is also possible for a node to have only one child; as in the case with the linear tree. An expression tree can be evaluated by applying the join operators at the root and the values obtained by evaluating the left and right sub trees that contain the relations keys. We evaluate each sub tree individually with postfix traversal by using reverse polish notation (RPN). The reverse polish notation (RPN) is a well-known method for the expression notification in a postfix way compared with the typical infix notation [9].

When comparing the reverse polish notation with algebraic notation, RPN has been found to achieve faster calculations [1]. Based on that, we will convert the multi-join queries into RPN expression in a postfix bottom-up manner instead of using the normal recursive multi-queries form. We construct the binary tree from the multi-join queries expression. Then, we use RPN to traverse it depending on the operand of the tree. Table1 presents an example of applying the RPN to SQL expression [4].

Figures2 and 3 show an example of a wide bushy tree and a linear tree of multi-join queries that uses RPN stacks. They present multi-join queries of six relations. At the beginning, we construct the binary trees where the leaves contain the operands of multi-join queries which are the relations keys. Join operation by default should contain two operands. Hence, the tree should at least contain two leaves and dynamically expands and shrinks depending on the number of join operations.

**Table 1.** Example of infix and RPN expressions [4].

| Infix expression | ((Department = 'Dep1')or(Department like 'Dep2%')) and ((Title = 't1') or (Title like 't2%')) |
|---|---|
| RPN expression | (Department = 'Dep1') (Department,like 'Dep2%') or (Title = 't1') (Title like 't2%') or and |

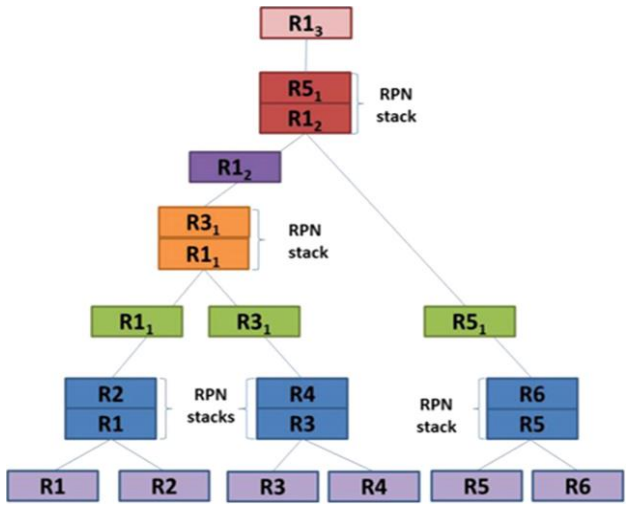

**Fig.2.** General (bushy) join sequence.

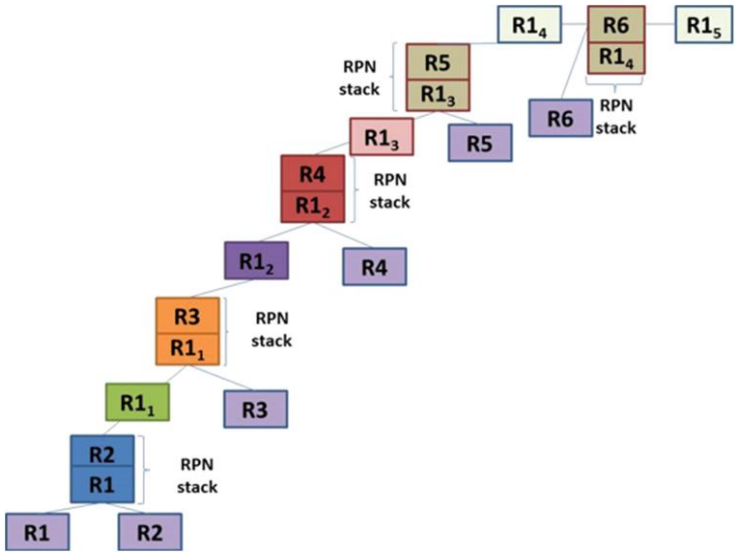

**Fig.3.** Sequential join sequence.

This solution has a minimum space complexity of $O(n)$, where n is the number of nodes of the tree. Figure 2 presents an example of wide bushy tree of six relations R1, R2, R3, R4, R5 and R6. The final join operation between $R1_2$ and $R5_1$ would be as following:

$$R1_3 \leftarrow (R1_2, R5_1) \bowtie \equiv ((((R1, R2) \bowtie)((R3, R4)) \bowtie)((R5, R6) \bowtie) \bowtie) \quad (2)$$

In linear tree as shown in Fig.3, we can notice that there is an extra joined relation $R1_5$ that needs one more join operation, which means more processing. The final joined relation can be obtained by:

$$R1_5 \leftarrow (R1_4, R6) \bowtie \equiv (((((R1, R2) \bowtie)(R3)) \bowtie)(R4) \bowtie)(R5) \bowtie)(R6) \bowtie) \quad (3)$$

Algorithm 1 presents the proposed technique by using subtree that adopts RPN.

---

**Algorithm 1.** Multi-join subtree that depends on RPN expression.

---

While nodes not empty

– **Read** *lc* **from** *subtree* – **Push**
  *lc→ st*

– **Read** *rc* of *lc* **from** *subtree* –
  **Push** *rc → st*

– **If** root is join operator

- **If** *st.lenght* < 2 and   IsNull(*subtree*.rightChild) **Then**

  **Report** an error

- **Else  pop**  *lc*,  **pop**  *rc*  **from**  *st*  jbt.leftChild  =(
  $R_{lc}, R_{rc}) \bowtie R_{lc.id} = R_{rc.id}$

---

## 4    Performance Evaluation

We conducted experiments on the proposed algorithm for concurrent join and compared it with linear sequence. We applied equi-joins which depends on equality (matching column values) where the primary keys of the tables are generated randomly from a predefined integer range. A two-phase strategy for multi-join queries is proposed in this paper.

The first phase studies a simple and cheap join algorithm, which is quicksort algorithm. Quicksort is a sorting-in place technique that applies divide and conquer algorithm. The advantage of this technique is that it is remarkably efficient on the average and it outperforms other sort-merge algorithms.

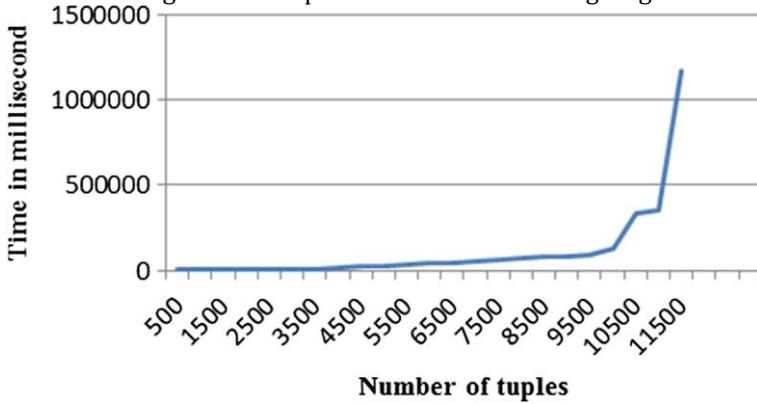

**Fig.4.** Execution time of join operation using RPN tree regarding number of tuples.

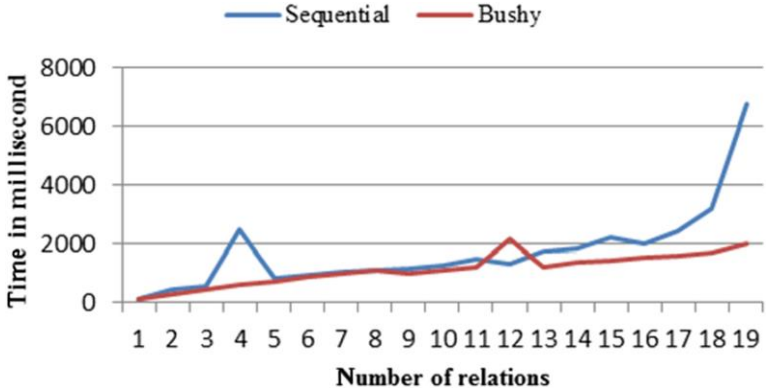

**Fig.5.** Execution time of join operation using sequential and bushy trees regarding number of relations.

The second phase applies reverse polish notation for reading the tree in bushy or linear manner. A relation, $R_i$, with $\|R_i\|$ records, that is populated with integer values have been used in the system. The system used for our evaluation was equipped with concurrent multi-threaded processor with 3MG cache and 6GB of system memory.

Figures 4 and 5 present the performance of executing the sequential and general trees regarding the number of tuples and tables. Figure 4 shows the performance of quicksort with RPN tree with 2 relations. Figure 5 shows the execution time of 100 tuples when the number of relations is increased. The bushy trees would be a better choice with larger number of tuples and relations. We notice that bushy trees still outperform the sequential trees.

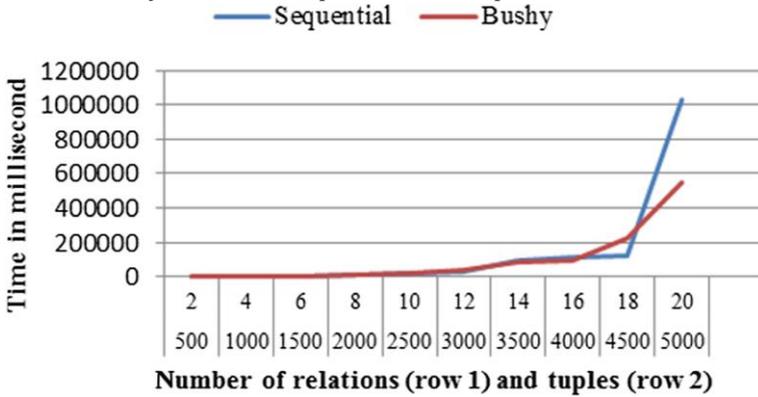

**Fig.6.** Execution time of join operation using sequential and bushy trees regarding number of relations and tuples.

**Table 2.** Other experiments on performance of sequential and bushy trees.

| # Tuples | # Relations | Time in millisecond | | # Tuples | # Relations | Time in millisecond | |
|---|---|---|---|---|---|---|---|
| | | Sequential | Bushy | | | Sequential | Bushy |
| 300 | 4 | 916 | 2071 | 900 | 10 | 5513 | 10808 |
| 500 | 6 | 1299 | 3520 | 1000 | 11 | 7108 | 13981 |
| 600 | 7 | 2101 | 3975 | 1100 | 12 | 5712 | 14488 |
| 700 | 8 | 2347 | 4221 | 1200 | 13 | 19916 | 27311 |
| 800 | 9 | 3372 | 9801 | 1400 | 15 | 19992 | 24937 |

Figure 6 also presents the execution time when increasing both tuples and relations. However, sequential could beat the bushy trees in case of having a small number of tuples and relations as shown in Table 2.

## 5    Conclusions

Join operation is still a vital step in most DBMS and the cost of queries is highly affected by this operation, hence, optimizing join operations leads to enhancing the DBMS. This paper presents a new methodology for sort-merge join by using RPN tree to perform multi join. We consider two factors in evaluating the performance: the number of tuples and the number of relations. Our approach was tested on both sequential and general trees. We concluded that the general tree outperforms the sequential tree in case of having a huge number of relations and tuples. In our future work, we plan to run the system on multi-core processors environment and also plan to expand this methodology to other queries operations.

**Acknowledgments.** We would like to thank Nada Alzahrani for her efforts during the progress of this research.